\documentclass [12pt, aps, epsf] {revtex4}
\usepackage {amssymb, amsbsy, epsf}

\begin {document}

\title {Use of Dynamical Undulator Mechanism to Produce Short Wavelength
Radiation in Volume FEL
(VFEL)}

\bigskip

\author {V.G.Baryshevsky}
\email [E-mail me at:] {bar@inp.minsk.by}
\author {K.G.Batrakov}
\email [E-mail me at:] {batrakov@inp.minsk.by}
\affiliation {Institute of
Nuclear Problems, Belarusian State University, 11 Bobruiskaya Str., Minsk
220050, Belarus.}
\begin{abstract}
VFEL lasing in system with dynamical undulator is described. In
this system radiation with longer
wavelength creates the undulator for
lasing on shorter wavelength. Two diffraction gratings with
different spatial periods form VFEL resonator. The grating with
longer period pumps the resonator by radiation of longer wavelength
to provide necessary amplitude of undulator field. The grating
with shorter period makes mode selection for radiation of shorter wavelength.
Lasing of such a system in terahertz frequency range is
discussed.
\end{abstract}
\maketitle

\section {\protect\bigskip Introduction}

The development of powerful electromagnetic generators with
frequency tuning in millimeter, sub-millimeter and terahertz
ranges using low-relativistic and non-relativistic
electron beams is quite perspective for numerous applications. In
such devices as TWT, BWT and orotrons generation of radiation with a
wavelength $ \lambda $ requires application of diffraction structures with period
$\sim \lambda \beta \gamma ^ {2} $. Only electrons passing near
the slowing structure at the distance
\begin {equation}
\delta \lesssim \delta _ {0} = \lambda \beta \gamma / (4\pi) \label {distance}
\end {equation}
interact
with electromagnetic wave effectively.  In (\ref{distance})
$ \beta=u/c $, $ \gamma $ is the Lorentz factor of electron.
 From (\ref{distance}) it follows,
 that generation of short-wave radiation for non-relativistic and
 low-relativistic electrons requires application
 of extremely thin and dense electron beams.
 Gyrotrons and undulator systems, in which
an electron beam interacts with an electromagnetic wave  in the
whole area occupied by magnetic field, has no such drawback.
Gyrotrons and cyclotron resonance facilities are used as radiation
sources in millimeter and sub-millimeter range, but their
operation requires magnetic field of about several tens of
kiloGauss ($\omega \sim \frac{eH}{mc}\gamma $). Applying undulator
systems for generation of short-wave radiation gives rise a
problem of manufacture of undulators with small period.  For
example, generation of radiation with the wavelength $ 1 $ mm at
the beam energy $E=800 $ KeV - $ 1 $ MeV requires to use undulator
with period $ 1 $ cm.  This is extremely complicated problem. Much
more rigid requirements arise in terahertz range. So even for
generation in longwave boundary of terahertz range the undulator
period should be $ \sim 3 $ mm (for $E=800 $ KeV). Nowadays such
requirement looks quite fantastic. For  radiation in millimeter
and sub-millimeter range it is possible to use mechanism of
Compton scattering of an electromagnetic wave by an electron beam.
Usually it is considered, that pumping wave emerges from outside.
But, radiation generated in a system also can play the role of a
pumping wave. If Q-factor of resonator is high enough, then the
radiation power is
  accumulated in resonator
  and can
act as undulator \cite{dynwig}. \ \

Such two-stage system (dynamic wiggler) can be used to soluve
several problems simultaneously:
1) it allows to obtain radiation of shorter wavelength
 ( $\lambda \sim \lambda _w/( 4\gamma^{2}) $, where $
\lambda _ {w} $ is the wavelength of a pumping wave. For example,
if electron beam has the energy $E \sim 800~KeV\div 1~MeV $ the
frequency increasef in $\sim \allowbreak 30$ times;

2) provides volume
character of interaction of an electron beam  with an electromagnetic wave
for non-relativistic beam  (lorentz
 factor $ \gamma \sim 1 $ ($u/c\ll 1 $)), for which dynamic wiggler can not
allow considerable increase of
radiation frequency.

For generation mechanisms based on slowing of electromagnetic wave
 (Cherenkov, Smith-Purcell, TWT, quasi-Cherenkov) electron beam interacts
with evanescent component  of electromagnetic wave (as it was noted
above), therefore, only  electrons, passing at the distances
(\ref{distance}) over slowing  structures, can radiate. Then, at the
first stage, the fraction of electron beam $\sim \delta _0/\Delta$
participates in creating of dynamic wiggler. All electrons
interact with the signal wave on the second stage.

Use of volume FEL principles extends advantages of the two-stage generation
scheme  and, in particular, allows  to tune smoothly the period of
dynamic wiggler by diffraction grating rotation. There is
possibility of smooth tuning of frequency of
both the pump wave and the signal wave
by smooth variation of geometrical parameters of volume
diffraction gratings, or by its rotation.
Moreover, use of VFEL
allows to create dynamic wiggler in large volume,
that represents a
great problem for static wiggler.

\begin{figure}[tbp]
\epsfxsize =8cm \centerline{\epsfbox{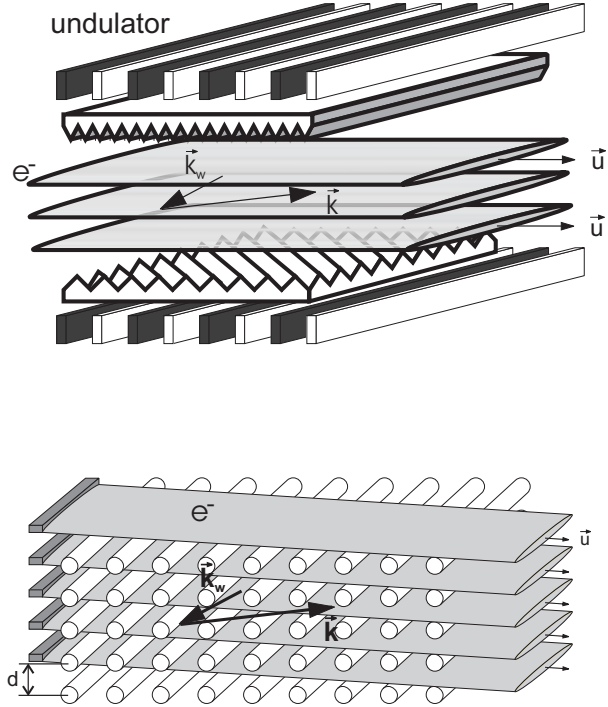}} 
\caption{ The scheme of dynamic wiggler } \label{wigg}
\end{figure}

This scheme can be realized by the use of resonator formed by
several diffraction gratings with different spatial periods
\cite{two_periods}. Smooth variation of geometrical parameters of
volume diffraction gratings  provides smooth variation of
frequency of pumping and signal waves. Two possible realizations
of dynamic wiggler on the basis of VFEL are represented on Figure
\ref{wigg}. Dynamic wiggler with undulator mechanism of the first
stage is depicted in the upper part of that figure. In this case
the the pumping wave with the vector $\vec k _w$ and frequency
$\omega _w \sim 2\gamma ^2 \frac{2\pi \beta}{d _u} $
  is formed by undulator with  period $d_u$. The wave with frequency
  $\omega \sim 4\gamma ^2 \omega_w$ is generated by the created dynamic wiggler
  during second stage. The distributed feedback is provided by two diffraction grating
  with different periods (one diffraction grating provides distributed
  feedback for pumping wave $\vec k_w$ and another
for signal one $\vec k$). The dynamic wiggler generated
by volume grating formed by threads is shown in the lower part
of Figure \ref{wigg}. The generation in VFEL
evolves in large volume with dimensions much exceeding
wavelength of radiation that increases electrical
endurance of resonator (due to  distribution of
electromagnetic power and electron beam over large volume). This
peculiarity of VFEL is extremely essential for generation of
powerful and super-powerful pulses of electromagnetic radiation.
The mode discrimination in such oversized system is carried out by
the multiwave dynamic diffraction \cite{basis}.

\section {Deriving of the basic expressions for electron beam instability
 in a pumping wave in conditions of multiwave diffraction.}

 There are two stages in the generation scheme proposed above:
a)generation of dynamic wiggler in a system with  two-dimensional
(three-dimensional) grating ; b) generation of radiation by an
electron beam  interacting with dynamic wiggler appeared on stage
a). Let's consider  stage a). The diffraction grating providing
non one-dimensional distributed feedback is used to evolve dynamic
wiggler. Smooth variation of diffraction geometry of this
diffraction grating  yields smooth variation of dynamic wiggler
parameters. The electromagnetic field of wiggler rises with time $
\tau_w\sim Q (\omega_w) /\omega_w $, where $ \omega_w $ is the
frequency of a pump wave. The Q-factor of resonator for frequency
$ \omega_w $ should be sufficient to create a magnetic field
amplitude about $ 100 $ Gs - $ 1 $ KGs. It follows from the energy
balance equation in resonator $ \frac {\omega_w} {Q} V\frac {H _
{m} ^ {2}} {8\pi} =W _ {0} $ ( $W _ {0} $ is the power of pump
wave formed by an electron beam ) that  $ Q =\frac {\omega_w } {W
_ {0}} V\frac {H _ {m} ^ {2}} {8\pi} $, where $V  $ is the cavity
volume, $H _ {m} $ is the amplitude of a magnetic field of dynamic
wiggler. When the pump field achieves necessary magnitude, the
stage b) begins. The created pump wave acts as undulator during
this stage. Non-one-dimensional distributed feedback can also be
realized for the signal wave. Dynamics of a signal electromagnetic
wave and an electron beam  in the system "volume diffraction
grating + pump electromagnetic wave " in this case is described by
equations
\begin {eqnarray}
DE-\omega ^ {2} \chi _ {1} E _ {1} -\omega ^ {2} \chi _ {2} E _ {2} -\omega ^
{2} \chi _ {3} E _ {3} -... &=&0 \nonumber \\ -\omega ^ {2} \chi _ {-1} E+D _
{1} E _ {1} -\omega ^ {2} \chi _ {2-1} E _ {2} -\omega
^ {2} \chi _ {3-1} E _ {3} -... &=&0 \label {system} \\ -\omega ^ {2} \chi _
{-2} E-\omega ^ {2} \chi _ {1-2} E _ {1} +D _ {2} E _ {2} -\omega
^ {2} \chi _ {3-2} E _ {3} -... &=&0, \nonumber
\end {eqnarray}
In (\ref {system}) $D _ {i} =k _ {\alpha} ^ {2} c ^ {2} -\omega ^
{2} \varepsilon + \chi _ {\alpha} ^ {(b)} $, $ \vec {k} _ {\alpha}
= \vec {k} + \vec {\tau} _ {\alpha} $ is the wave vector of
diffracted wave in the field of volume grating , $ \chi _ {\alpha} ^
{(b)} $ is the part of dielectric susceptibility corresponding to
interaction of an electron beam with radiation
\begin{eqnarray}
\chi _{\alpha }^{(b)}=\frac{q^{(w)}}{\left\{ \omega -\vec{k}_{\alpha
}\vec{v}_{w}\mp (\omega ^{(w)}-\vec{k}_{\beta }^{(w)}\vec{v}_{w})\right\} ^{2}}
&&\text{ ''cold''\ beam limit,}  \nonumber \\
\chi _{\alpha }^{(b)}=-i\sqrt{\pi }\frac{q^{(w)}}{\delta _{\alpha
}^{2}}x_{\alpha }^{t}\exp [-(x_{\alpha }^{t})^{2}] &\ \ \ \ \ &\text{''hot''\ beam
limit.}  \nonumber
\end{eqnarray}\begin{eqnarray*}
q^{(w)} &=&\frac{a_{w}^{2}}{4\gamma ^{3}}\frac{\omega
_{L}^{2}}{(k_{w}v)^{2}}\left\{ \left[ \frac{\vec{u}\vec{e}}{c(k_{w}v)}(\omega
_{w}\vec{u}-\vec{k}_{w}c^{2})(\vec{k}-\vec{k}_{w})-(\vec{k}\vec{e}_{w})c\right]
\frac{\vec{u}\vec{e}}{c}-(\vec{k}_{w}\vec{e})(\vec{u}\vec{e}_{w})-(\vec{e}\vec{e}%
_{w})(k_{w}v)\right\} ^{2}\cdot \\
&&\left\{ (\vec{k}-\vec{k}_{w})^{2}c^{2}-(\omega -\omega _{w})^{2}\right\}
\end{eqnarray*}%
$ \vec {k}, \omega, \vec {e} \ $and $ \vec {k} _ {w}, \omega _
{w}, \vec {e} _ {w} $ are the wave vectors, frequencies and
polarization vectors of  both the  signal  and pumping waves
accordingly, $v = (c, \vec {u}) $, $k _ {w} = (\omega _ {w}, \vec
{k} _ {w}) $, $a _ {w} = \frac {eH _ {w}} {mc\omega _ {w}} $.
 Dispersion equation
corresponding to (\ref {system}) has the following schematic form
\begin {equation}
D ^ {(n)} = -\chi _ {\alpha} ^ {(b)} q D ^ {(n-1)} \label {disp}
\end {equation}
the term $D ^ {(n)} $ in  the left-hand side (\ref {disp}) corresponds
to $n  $-wave Bragg dynamical diffraction \cite{chang} (equation
$D ^ {(n)} =0 $ is the  dispersion equation defining diffraction modes
in $n $-wave case). For the limit of "cold" electron beam  the
equation (\ref {disp}) has $2n+2 $ solutions. In this case field
in a system is represented as
\begin{equation}
\vec{E}(\vec{r};\omega )=\sum \vec{E}^{(j)}s_{i}^{(j)}\exp
\{i(\vec{k}+\vec{\tau}_{i})\vec{r}\}.  \label{field}
\end{equation}
 the summation over index  $i=1\div n $ ($n $ is
the number of strongly coupled  waves) and over index $j=1\div
2n+2 $ ($2n+2 $ is the number of solutions of the dispersion
equation (\ref {disp})) is fulfilled in (\ref {field}), $ \vec {E}
^ {(j)} $ is the amplitude of $j^{th}$ mode.
According to solutions of
(\ref {disp}), $s _ {i} ^ {(j)} $ is the coupling coefficient of
diffraction waves in $j ^ {th} $ mode.

The $2n+2 $ amplitudes $ \vec {E} ^ {(j)} $ are defined by
the boundary conditions. The
continuity of current densities, charge densities and transverse
components of fields on boundaries are used:
\begin {eqnarray}
\sum _ {j} (\vec {f} ^ {(\alpha)} \vec {E} ^ {(j)}) s _ {1} ^ {(j)} e _ {i} ^
{(j)} &=&a_{1}^{(\alpha)} \label {boundary} \\
\sum _ {j} (\vec {f} ^ {(\alpha)} \vec {E} ^ {(j)}) s _ {2} ^ {(j)} e ^ {(j)}
 &=&a_{2}^{(\alpha)} \nonumber \\ &&... \nonumber \\
 \sum_{j}\frac{(\vec{u}\vec{E}^{(j)})}{\delta _ {j}} &=&0 \nonumber \\
\sum_{j}\frac{(\vec{u}\vec{E}^{(j)})}{\delta _ {j} ^ {2}} &=&0. \nonumber
\end{eqnarray}
In (\ref {boundary}) $ \vec {f} ^ {(\alpha)} (\alpha =1\div 2) $
are two vectors of polarization, $ \delta _ {j} = \frac {\omega -k
_ {jz} u\mp \Omega} {\omega} $ is the detuning parameter,
\[
e_{i}^{(j)}=\left\{
\begin{array}{c}
1\ \text{if }k_{jz}+\tau _{iz}>0\text{ } \\
\exp \{ik_{jz}L\}\ \text{if }k_{jz}+\tau
_{iz}<0,\end{array}\right.
\]$a_{i}^{(\alpha )}$ are the amplitudes of waves
emerging in a system from the outside.

The requirement of equality of determinant of (\ref {boundary}) to
zero corresponds to excitation of a system in the absence of incident
waves.
Respective expression is named the generation threshold
equation.

The expansion (\ref {field}) of a field is valid for oversized
system, when
transverse dimensions
exceed
radiation  wavelength.
If ttransverse dimensions are comparable with a
wavelength, the expansion over plane waves is replaced by the expansion
over eigenmodes of waveguide. The general form of boundary
conditions (\ref {boundary}) and equation of generation does not
vary, all peculiarities of a system are contained in the form of
coupling coefficients  between waves (modes) $s _ {i} ^ {(j)} $.

In the range of degeneration of $n $ roots of the dispersion equation
(when $n+1 $ solutions of the equation $D ^ {(m)} =0 $ coincide at
$m\geq n+1 $ ) solution of the generation threshold equation
has the form \cite{basis}:

\begin {equation}
\frac {1} {\gamma ^ {3}} \left (\frac {\omega _ {L}} {\omega} \right) ^ {2} a _
{w} ^ {2} k ^ {3} L _ {\ast} ^ {3} = \frac {a _ {n}} {(k |\chi |L _ {\ast}) ^
{2n}} +b _ {n} k\chi " L _ {\ast}, \label {threshold}
\end {equation}
here $k =\omega /c $,  $L _ {\ast} $ is the
length of interaction range of an electron beam  with
electromagnetic radiation, $ \chi " $ is the imaginary part of
dielectric susceptibility describing absorption of radiation,
$a _ {n} $, $b _ {n} $ are the magnitudes depending on geometrical
parameters of a system (except for $L _ {\ast} $). The equality
(\ref{threshold}) has an obvious physical meaning: in the
left-hand side there is a term describing generation of radiation
by an electron beam, and in the right-hand side there are the terms
describing losses of the radiation on boundaries (the first term
in a right-hand side of (\ref {threshold})) and losses determined by
radiation absorption (the second term in a right-hand side of
(\ref{threshold})). One of advantages of VFEL  with $n-$wave
distributed feedback is the possibility of  sharp decreasing of
boundary losses.  It follows from the fact that in conditions
of  dynamical diffraction  $k |\chi |L _ {\ast } \gg 1$ and
therefore the first term in the right-hand side of
(\ref{threshold}) decreases with the number of wave $n$ increase.
Let us note that the system can work as generator without  dynamic
diffraction (in the one-wave case). Such situation is realized if
a  wave has a group velocity directed opposite to  electron beam
velocity. In this case  (\ref {boundary}) looks like

\begin{eqnarray}
c ^{(1)}\exp \{ik\delta _{1}L\}+c ^{(2)})\exp \{ik\delta
_{2}L\}+c ^{(3)})\exp \{ik\delta _{3}L\} &=&0  \label{one_wave} \\
\frac{c ^{(1)}}{\delta _{1}}+\frac{с ^{(2)}}{\delta
_{2}}+\frac{c ^{(3)}}{\delta _{3}} &=&0  \nonumber \\
\frac{c ^{(1)}}{\delta _{1}^{2}}+\frac{c ^{(2)}}{\delta
_{2}^{2}}+\frac{c ^{(3)}}{\delta _{3}^{2}} &=&0.  \nonumber
\end{eqnarray}
The generation equation corresponding to (\ref {one_wave}) has the
form

\begin{equation}
\delta _{1}^{2}(\delta _{2}-\delta _{3})\exp \{ik\delta
_{1}L\}-\delta _{2}^{2}(\delta _{1}-\delta _{3})\exp \{ik\delta
_{2}L\}+\delta _{3}^{2}(\delta _{1}-\delta _{2})\exp \{ik\delta
_{3}L\}=0.  \label{one_t}
\end{equation}
It  follows from the analysis of (\ref {one_t}), however, that
generation in this case takes place only in the regime  of strong
amplification ($k Im\delta L_{\ast }\gtrsim 1$). Besides, the
frequency of a signal wave has the same order or less then the
frequency of a pumping wave in this case. Therefore, such branch
of generation does not represent interest.

\section{Discussion}
\begin{figure}[tbp]
\epsfxsize =8cm \centerline{\epsfbox{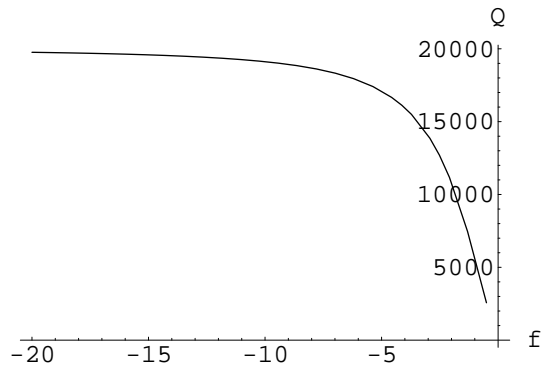}} 
\caption{ Calculated dependence of Q factor  on diffraction
asymmetry factor $\beta$ } \label{Fig.2}
\end{figure}

Let's write  the condition of  synchronism between electron beam and
signal wave in VFEL at the presence of pumping wave created at the
first stage:
  $ \omega -\vec {k} \vec {v} _ {w} = $ $ \Omega _
{w} $. Here $ \Omega _ {w} = \omega _ {w} + \vec {v} _ {w} \vec
{k} _ {w} $. If generated system is oversized with respect to
pumping wave (transverse dimensions essentially exceed the pumping
wavelength $D \gg \lambda$), otherwise $ \Omega _ {w} = \omega _
{w} +hv _ {wz} $, where $h $ is waveguide eigenvalue for the
pumping wave. So the frequency of signal wave is approximately
equal to

\begin {equation}
\omega = \frac {2\omega
_{w}(\vec{\tau}_{1},..\vec{\tau}_{n},\vec{n}_{u},S)}{1-n(\omega) \beta \cos
(\Theta)}. \label {freq}
\end {equation}
Dependence of pumping wave frequency of multiwave diffraction
geometry ( $ \vec {\tau} _ {1}, ..\vec {\tau} _ {n} $) and
geometry of the resonator $S $\ (if system is not oversized  with
respect to pumping wave, in opposite case dependence on $S $
misses) is specially noted in (\ref {freq}). Thus using VFEL with
dynamic wiggler allows to control the period of wiggler and the
frequency of a signal wave by either diffraction grating
rotation or change of velocity  direction.  Smooth change of
VFEL geometry also varies
 the Q-factor  and, therefore, varies the generation  efficiency.
 For example
 dependence of Q  on  diffraction asymmetry factor $\eta = \gamma_0 /\gamma_1$
 for a two-wave case is shown
  on Fig.\ref{Fig.2}.
 The signal wave can also be in condition of dynamical diffraction.
Bragg synchronism in this case is provided by  diffraction grating
with lesser period (the wavelength of signal  wave is less then
the wavelength of pumping wave) \cite{two_periods}. For radiation
angle $ \Theta =0 $ $ \omega \sim 4\gamma^ {2} \omega _ {w} $
(\ref {freq}), thus, even moderately relativistic electron beams
with $E\sim 1 $ MeV gives the multiplications of frequency is of
order $4\times (1+1000\div 511) ^ {2} \sim \allowbreak 35 $. If
the first stage is based on undulator mechanism with undulator
period $\sim 8$ cm,
 then the wavelength of pumping
wave $\lambda_w \sim 1$ cm. It allows to generate the signal wave
in terahertz range.

Thus, it is shown that
\begin {itemize}
\item The principles of VFEL can be used for creating of dynamic
wiggler
 with varied period in large volume;

\item Two-stage scheme of generation can be used for
generation by low-relativistic beams
 in terahertz
frequency range ;

\item Two-stage scheme of generation combined with volume
distributed feedback gives possibility to create powerful
generators with wide electron beams (or system of beams).

\end {itemize}

\end {document}